\documentclass[10pt,final,journal,twocolumn]{IEEEtran}
\ifCLASSINFOpdf
\else
\fi
\usepackage{cite}
\usepackage{multirow}
\usepackage{longtable}
\usepackage{amsfonts}
\usepackage[cmex10]{amsmath}
\usepackage{amsthm}
\usepackage{algorithmic}
\usepackage{stfloats}
\usepackage{multicol,multienum}
\usepackage[dvips]{graphicx}

\begin{document}

\title{A Game Theoretic Perspective on Self-organizing Optimization for Cognitive Small Cells}

%\author{
%Yuhua~Xu,~\IEEEmembership{Member,~IEEE,}
%  Jinlong~Wang,~\IEEEmembership{Senior Member,~IEEE,}
%        Qihui~Wu,~\IEEEmembership{Senior Member,~IEEE,}
%          Zhiyong Du,
%        Liang Shen, and
%        Alagan~Anpalagan,~\IEEEmembership{Senior Member,~IEEE}% <-this % stops a space

\author{
Yuhua~Xu,
  Jinlong~Wang,
        Qihui~Wu,
          Zhiyong Du,
        Liang Shen and
        Alagan~Anpalagan

\thanks{This work was supported by the National Science Foundation of China
under Grant No. 61401508 and No. 61172062.}
\thanks{Yuhua~Xu, Jinlong~Wang, Qihui~Wu, Zhiyong Du and Liang Shen are with the College of Communications Engineering, PLA University of Science and Technology, Nanjing, China (e-mail: yuhuaenator@gmail.com, wjl543@sina.com, wuqihui2014@sina.com, duzhiyong2010@gmail.com, ShenLiang671104@sina.com ).}
\thanks{Alagan~Anpalagan is with the Department of Electrical and Computer Engineering, Ryerson University, Toronto, Canada (e-mail: alagan@ee.ryerson.ca).}
}% <-this


\IEEEpeerreviewmaketitle
\maketitle
\begin{abstract}
In this article, we investigate self-organizing optimization for cognitive small cells (CSCs),   which have the ability to sense the environment, learn from historical information,  make intelligent decisions, and  adjust their operational parameters.
By exploring the inherent features, some fundamental challenges  for self-organizing optimization in CSCs are presented and discussed. Specifically, the dense and random deployment of CSCs brings about some new challenges in terms of scalability and adaptation; furthermore, the uncertain, dynamic and incomplete information constraints also impose some new challenges in terms of convergence and robustness. For providing better service to the users and improving the resource utilization, four requirements for self-organizing optimization in CSCs are  presented and discussed. Following the attractive fact that the decisions  in game-theoretic models are  exactly coincident with those in self-organizing optimization, i.e., distributed and autonomous, we establish a framework of game-theoretic solutions for self-organizing optimization in CSCs, and  propose some featured game models. Specifically, their basic models are presented,  some examples are discussed and  future research directions are given.
\end{abstract}


% \fi
\IEEEpeerreviewmaketitle

\section{Introduction}
\IEEEPARstart{S}{mall} cells have been regarded as a promising approach to meet the  increasing demand of cellular network capacity. In comparison to macro-cells, low-cost small cells operating with low-power and short range offer a significant capacity gain due to spatial reuse of spectrum. Researchers in the community have realized that enabling cognitive ability into small cells,  which is  referred to as cognitive small cells (CSCs) \cite{small_cell2}, would further improve the resource utilization. Similar to cognitive radio, CSCs are able to sense the environment, learn from historical information,  make intelligent decisions, and  adjust their operational parameters.

It is expected that small cells are to be \emph{densely }deployed in the near future. Furthermore, small cells may be deployed by the mobile operators, the enterprises or households, which means that they would operate in a self-organized, dynamic and distributed manner. Thus,  resource optimization problems for small cells, e.g., spectrum sharing, carrier selection and power control, interference management, and offloading mechanism, can not be solved in a centralized manner since it results in heavy communication   overhead and can not adapt to  dynamic environment. As a result,
it is important and timely to develop self-organizing optimization approaches for CSCs.

In this article, by exploring the inherent features of CSCs, we first discuss and analyze some fundamental  challenges  and requirements for self-organizing optimization  in CSCs. Following the attractive advantages of game-theoretic models for self-organizing optimization, we propose some featured game-theoretic solutions. It should be pointed out that there are also some other useful approaches for self-organizing optimization in distributed wireless networks, e.g., the swarm intelligence inspired evolutionary algorithms \cite{self_organization}. The reasons for using game-theoretic solutions  are: i) the interactions among multiple decision-makers can be well modeled and analyzed, and ii) the outcome of the game  is predicable and hence the system performance can be improved by manipulating the utility function and the action update rule of each decision-maker.

Game-theoretic models have been investigated extensively in wireless communications, and there are some preliminary game-theoretic solutions for CSCs, e.g., reinforcement learning with
logit equilibrium for power control \cite{related1}, a hierarchical dynamic game approach for spectrum sharing and service selection \cite{Stackelberg_game},  and evolutionary game for self-organized resource allocation \cite{related2}. The presented models in this article mainly address  the inherent features, fundamental requirements and challenges  of CSCs  and hence differentiate significantly from previous ones seen in the literature. In fact, the main objective of this article is to propose and discuss the featured game-theoretic models suitable for CSCs.

The rest of this article is organized as follows. In Section II,  the cognition functionality for CSCs is presented. In Section III, some fundamental challenges and requirements for  CSCs are discussed. In Section IV, some featured game-theoretic models for self-organizing optimization in CSCs are presented, and  future research directions are given.
Finally, we provide concluding remark in Section IV.

\section{Cognition and Self-organizing Optimization for Cognitive Small Cells}

We first present the cognition functionality for CSCs, which is the base for self-organizing optimization.
The cognition functionality in cognitive radio is mainly concerned with acquiring spectrum availability information, i.e.,  sensing and identifying spectrum opportunities in time, frequency and space domains. To capture the complex environment and network state, the cognition functionality in CSCs is extended  to explore  multi-dimensional information. Such multi-dimensional information is referred to as\emph{ contextual  information}, which is used to  identify an object of interest.

 \begin{figure*}[bt!]
  \center
  \includegraphics[width=6.0in]{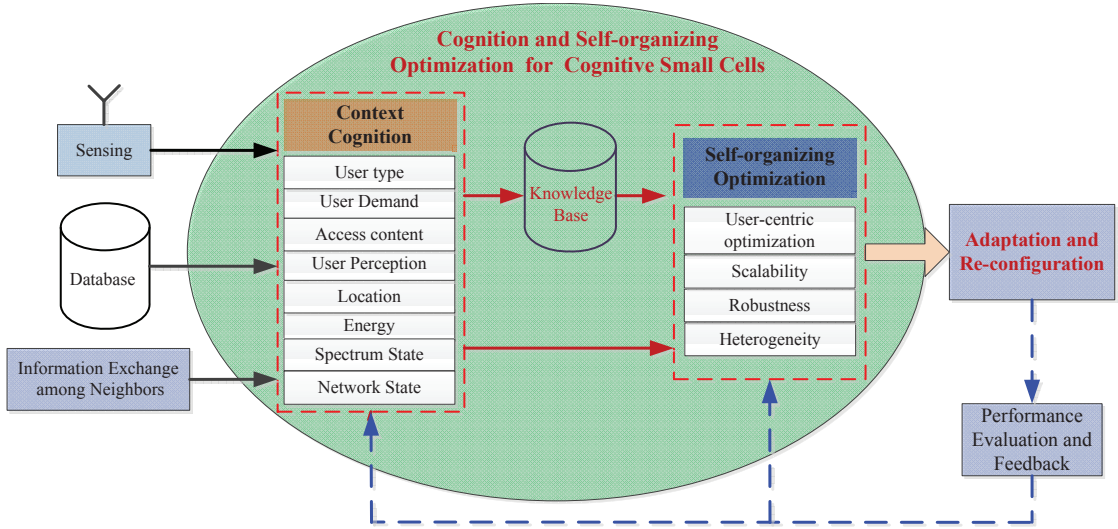}\\
  \caption{The paradigm of cognition functionality and self-organizing optimization in  CSCs.}
  \label{fig:paradgim}
\end{figure*}

 As illustrated in Fig. \ref{fig:paradgim}, the target contextual information includes user type, user demand, access content, user perception, location, energy, network state and spectrum state. For presentation, we briefly illustrate the above contextual  information.  The user type represents the hardware category i.e., tablet, phone or a laptop. Access content is related to specific application, e.g., browsing a breaking news or downloading an APP.  User perception is related to the service quality experienced by the user, while location is where the small cell is, e.g., indoor or outdoor. Energy is related to the power for wireless transmission and cooling. Network state is related to the current network deployment of small cells and spectrum state is related to the spectrum availability. Technically, the above contextual  information has great impact on the resource allocation schemes, which will be discussed later.

 As shown in Fig.\ref{fig:paradgim},  the contextual  information in CSCs can be obtained by the following three methods: sensing, database, and information exchange among neighbors.
i) Sensing: with the help of cognitive radio technologies, CSCs perform sensing to obtain  useful information. For example, the spectrum occupancy state can be obtained by energy detection or feature detection (e.g., pilot, modulation type, cyclic prefixes, and cyclostationarity). Sensing is  real-time but consumes resources including time, energy and bandwidth.
 ii) Database: database approach is a powerful tool to provide useful information for CSCs. For example, spectrum database has recently been developed to provide spectrum occupancy state for a particular region, through which the CSCs can inquire the location-dependent spectrum availability information. Compared with sensing, database approach is more efficient but is not real-time. iii) Information exchange among neighbors: since the CSCs are connected to the core network via cable or optical fiber, information exchange among  CSCs are feasible. However, it is to be noted  that local information exchange between neighbors is more desirable since global information exchange leads to heavy communication and signalling overhead.

We argue that the term ``cognitive" in CSCs is not limited to  observing the environment and acquiring contextual  information. Instead, it should include having high-level intelligence. To achieve such high-level intelligence for CSCs, the most promising way is to realize knowledge discovery from the contextual  information. Generally speaking, knowledge is a broad concept including  general principles and natural laws. Taking the spectrum dynamics as an example, the probability $\theta$ that a particular band is occupied by the macro-cells during  01:00 AM to  04:00 AM is very small, e.g., $\theta=0.05$, is viewed as the knowledge in CSCs.
Based on the contextual  information, the CSCs can build  knowledge base which contains useful knowledge for self-organizing optimization.

\section{ Challenges and Requirements for Self-organizing Optimization in Cognitive Small Cells}
In this section, by exploring the inherent features of CSCs, we briefly discuss  some fundamental challenges and requirements for self-organizing optimization in CSCs.

\subsection{Technical challenges}
The technical challenges for self-organizing optimization in CSCs are discussed from the perspectives of network deployment and information constraints respectively.

First, from the perspective of network deployment of CSCs, two challenges arise in the following two aspects:
\begin{itemize}
  \item \textbf{Scalability}. With the increase in the  number of  small cells, how would the  self-organizing optimization solutions scale up? This is the first basic issue of dense deployment in CSCs. In addition, since the decisions of CSCs are interactive, addressing the complicated interactions among the densely deployed CSCs is another important issue.
  \item \textbf{Random  deployment}.  The small cells are deployed by different entities, e.g., mobile operators, the enterprise or households. In addition, they may turn to be inactive if there is no serving client. As a result, the deployment of small cells is always random and dynamic. Thus, it is important for the self-organizing optimization solutions to behave  in random and dynamic environment.
\end{itemize}

Secondly, it is known that information is key to optimization problems, and the challenges related to information arising in the CSCs are listed  below:
\begin{itemize}
  \item {\textbf{Uncertain}}: the observed information may  not be the same with the true information. A well-known example is that the sensed spectrum states are  always imperfect due to the corruption of   noise.
  \item {\textbf{Dynamic}}: the observed information is time-varying the dynamic changes are not determinate. For example, the network and spectrum states may change from time to time, and the set of active CSCs may  also change from time to time. Furthermore, the network and spectrum states in each decision period are random, the set of active CSCs are random, and their demands are also random.

  \item {\textbf{Incomplete}}: due to the constraints in  hardware and resource consumption, each CSC only  has partial information about the environment; furthermore, it only has information of its neighboring CSCs (in some extreme scenarios, it has no information of others). In addition, a CSC does not know the total number of small cells in the systems, not to mention the active ones.
\end{itemize}

Due to the above technical challenges, it is seen that the task of  resource optimization in CSCs is hard to solve even in a centralized manner, not to mention in a distributed and self-organizing manner.

\subsection{Requirements for self-organizing optimization}
We list some fundamental requirements for self-organizing optimization in CSCs. Specifically, these requirements are  for user service, network deployment (architecture), and optimization methodology. As shown in Fig. \ref{fig:paradgim}, based on the contextual  information and knowledge, some self-organizing optimization approaches can be applied for resource allocation in CSCs. By employing their inherent features, we discuss some featured requirements of self-organizing optimization in CSCs, which mainly include user-centric optimization, scalability, robustness and heterogeneity.

First, it should shift from throughput-oriented optimization to user-centric optimization.  Traditionally,  resource  optimization schemes in wireless systems are throughput-oriented, with the  objective to maximize throughput/capacity or minimize  delay. However, it is now realized that throughput-oriented schemes can not provide satisfactory service for the users. In future mobile communication systems, there is an increasing trend to develop user-centric optimization schemes rather than throughput-oriented schemes. The underlying reasons are twofold: (i) eventually, the purpose of (wireless) communication is  to serve end users. Thus, the contextual  information of the users, e.g., their locations, demands,  access contents and energy, should be taken into account not only in high-layers but also in PHY and MAC layers for optimization, (ii) it is realized that mobile (cellular) systems have migrated towards data and Internet services. In particular, multimedia service delivery through cellular systems, e.g., watching online video, is becoming common. For this kind of service, people may not care about the specific volume of allocated resources, but sensitively react to the perceived service quality, which is known as quality of experience (QoE) \cite{QoE}. It means that user perception should also be taken into account in self-organizing optimization.

Secondly,  it should admit scalability and  address network density. As stated before, it is expected that the CSCs are densely deployed with large numbers. A consequence is that the resource optimization for dense deployment  is completely different than that in sparse environment. Thus, the self-organized optimization schemes should scale up in densed CSCs. In addition, density creates  congestion and interference among CSCs, which implies that efficient congestion control and interference mitigation approaches should be developed.

Thirdly, it should be robust to the dynamic environment. As discussed before, there are several random and dynamic factors in CSCs, e.g., the spectrum availability is dynamic, the CSCs switch between active and inactive randomly. Moreover, the observed information may be corrupted by noise. Thus, the self-organizing optimization solutions should be robust to address the randomness, dynamics, and uncertainty in CSCs.

Last,  it should address the hierarchical decision-making in CSCs. There are always heterogeneous cells with overlapping coverage in future wireless systems, i.e., macro-cells and small cells. In such hierarchical networks, the cells in different layers have different priority and utility functions. That is, it involves heterogeneous decision-makers. However, traditional  self-organizing optimization solutions are mainly for homogeneous decision-makers. Thus, it is important to develop new hierarchical self-organizing solutions for CSCs.

\section{Game-theoretic  Self-organizing Optimization for Cognitive Small Cells}
Game theory \cite{game_book} is an applied mathematic tool to model and analyze  mutual interactions in multiuser decision systems. Generally, a game model consists of a set of players, a set of available actions of each player, and a utility function that maps the action profiles of the all the players into a real value. There are two major branches of game-theoretic models:  non-cooperative games and cooperative games. From a high-level perspective of comparison, players in a non-cooperative game make rational decisions to maximize their individual utility functions, while players in a cooperative game are grouped together according to an enforceable agreement for payoff allocation. In a non-cooperative game, the commonly used solution concepts are Nash equilibrium (NE) and correlated equilibrium.

Researchers began to apply game-theoretic models into wireless communications a decade ago; nowadays, it has been regarded as a powerful tool for wireless resource allocation  optimization, e.g., power control, spectrum access, network selection, spectrum auction and trading, and incentive mechanism design. The decisions of the players in (non-cooperative) game-theoretic models are distributed and autonomous, which is an exact coincidence with those in self-organizing optimization. Thus, game-theoretic approach is  important to achieve self-organizing optimization in CSCs \cite{game_csc_book}.

 \begin{figure}[bt!]
  \center
  \includegraphics[width=3.2in]{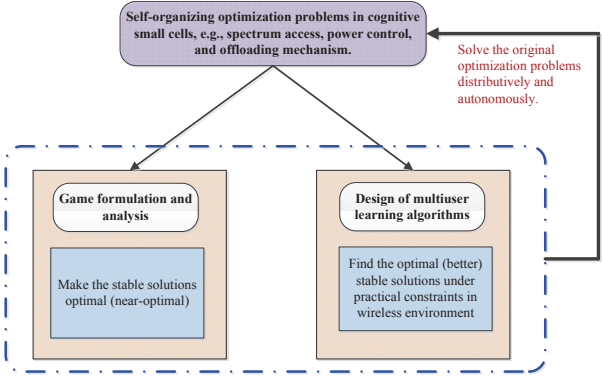}\\
  \caption{The proposed framework of game-theoretic solutions for self-organizing optimization in CSCs.}
  \label{fig:game_optimization}
\end{figure}

\subsection{Framework of game-theoretic  self-organizing optimization }
To cope with the technical  challenges in CSCs, i.e., dense and dynamic deployment, and uncertain, dynamic and incomplete information constraints, we  propose a framework of game-theoretic self-optimizing optimization, which is  shown in Fig. \ref{fig:game_optimization}. It is noted that there are two key steps: i) game formulation and analysis, and ii) design of multiuser learning algorithm. In comparison, game formulation and analysis belongs to theoretical investigation while developing learning algorithms belongs to the field of algorithm methodology. On one hand, the stable solutions are the inherent properties of game-theoretic models, and  not relevant to the learning algorithms. On the other hand, except for the utility function in game-theoretic models, the uncertain, dynamic and incomplete information constraints have great impact on the convergence and performance of learning algorithms. With the two steps carefully designed, self-organizing optimization for the original wireless optimization problems can be achieved.

\subsubsection{Game formulation and analysis}
For game formulation, one needs to first identify the player and available action set, and define suitable utility functions for the players. For CSCs, the player may be a single entity, e.g., small base station or the user equipment, or a collection of multiple entities, e.g., a cluster consisting of multiple nearby small cells. The available action set can be regarded as a combination of multiple optimization variables. Defining utility function is key to game formulation since it eventually determines the properties and performance of the game-theoretic models.

The most efficient game-theoretic model used in wireless networks is potential game \cite{Monderer96}, in which there is a potential function such that  the change in the utility function caused by the unilateral action change of an arbitrary player has the same trend with that in the potential function, i.e., both increasing or decreasing. Potential game has at least one pure strategy NE and all NE points are global or local maxima of the potential function. Thus,  the NE solutions are desirable if the potential function is related to the original optimization objective. Furthermore, to ensure that the stable solutions of game-theoretic models are optimal (near-optimal), an another efficient method is to define the utility function as the received payoff minus cost of using the amount of particular resource.

\subsubsection{Design of multiuser learning algorithms }

 Identifying the stable solutions of game-theoretic models is one thing, but finding them is another  different thing. This issue, however, was underestimated in previous studies. In pure game theory, players can monitor the environment and other players, which means that they have perfect information about the actions and payoffs of other players. As discussed above, this assumption does not hold in CSCs. With the cognition functionality of CSCs, players need to observe the results of multiuser interactions, e.g., interference, collision and competition, learn useful information from the limited feedback, and then adjust their behaviors towards some desirable solutions. In the context of game optimization, the objective of multiuser learning is to converge to a stable solution with good performance.

 Denote $a_n(k)$ as the action of player $n$ in the $k$th iteration, and $a_{-n}(k)$ as the action profile of all other players except $n$.
 Due to the interactions (interference, congestion or competition) among players, the received payoff $r_n(k)$ of each player is jointly determined by the action profile of all players,  and it may be deterministic or random. Generally, the players update their actions based on the current action-payoff information $\big\{ {{a}_{n}}(k),{{a}_{-n}}(k);{{r}_{n}}(k),{{r}_{-n}}(k) \big\}$. Thus, the  system evolution can be described as $\left\{ {{a}_{n}}(k),{{a}_{-n}}(k) \right\}\to \left\{ {{r}_{n}}(k),{{r}_{-n}}(k) \right\}\to \left\{ {{a}_{n}}(k+1),{{a}_{-n}}(k+1) \right\}$, and the objective is to converge to a stable action profile that maximizes the system utility.

The uncertain, dynamic and incomplete information constraints in CSCs may pose some new challenges. Specifically,
i) a player does not know the information about all other players, i.e., $a_{-n}(k)$ and $r_{-n}(k)$ are unknown,  ii) the received payoff $r_n(k)$ may be random and time-varying. Thus, the update rule needs to be carefully designed to guarantee the convergence towards desirable solutions. When local information among neighboring players is available, it is desirable to develop partially uncoupled learning algorithms based on the partial action-payoff information $\big\{{{a}_{n}}(k),a_{J_n}(k);{{r}_{n}}(k), r_{J_n}(k) \big\}$, where $a_{J_n}(k)$ and $r_{J_n}(k)$ are the action-payoff information of the neighboring players. In some extreme scenarios with no information exchange available, one needs to develop fully uncoupled learning algorithms based on the individual action-payoff information $\big\{{{a}_{n}}(k),{{r}_{n}}(k) \big\}$. There are some useful partially coupled learning algorithms, e.g., local altruistic behavior with spatial adaptive play \cite{graphical_game}, and fully uncoupled learning algorithms, e.g., stochastic learning automata \cite{SLA}, that can be applied for self-organizing optimization in CSCs.

\subsection{Some featured game-theoretic models for CSCs }
In methodology, previous game-theoretic models in wireless communications can  be applied to CSCs. However,  most previous game-theoretic solutions mainly focused on analyzing the properties of game-theoretic models in ideal scenarios, and did not take into account the  challenges and requirements in CSCs. In this subsection, we propose some featured game modes for CSCs. Due to the low-power, the transmission of a small cell only affects its neighbors; as a result, graphical games \cite{graphical_game} are appropriate for small cell networks.

\emph{1) Demand-aware game:} Most existing resource optimization approaches mainly focused on maximizing the resource utilization, while ignoring the actual demand of the users. In future CSCs, demand-aware design and decision is more desirable. To include the user demand into the resource optimization problems, a useful method is to map the allocated resource to the user satisfaction utility. Specifically, denote  user $n$'s demand as $d_n$ and the allocated  resource as $r_n$, then its satisfaction utility can be expressed as $s_n(r_n, d_n)$.

Generally, there are two kinds of satisfaction functions in the literature: i) the linear satisfaction: the satisfactory utility is determined by $\frac{r_n}{d_n}$ if $r_n\le d_n$, and is equal to one otherwise, and ii) the sigmoid  satisfaction: the satisfactory utility function is generally determined by $\frac{1}{1+e^{-c(r_n-d_n)}}$, where  $c$ is used to adjust the slope of the satisfaction utility curve around the user demand $d_n$ with different types of traffic. In particular,  real-time traffic such as online video are sensitive to acquired resources and require strict performance requirements, which corresponds to large value of $c$, while non-real-time traffic such as E-mail or file transfer are less sensitive, which corresponds to small value of $c$. The linear and sigmoid satisfaction functions have been well investigated in previous game-theoretic wireless resource optimization problems. As the satisfaction utility function is strictly increasing, each user  proceeds to compete for wireless resource even if the obtained resource is larger than the demand, which would decrease the satisfaction of others. However, this drawback has not addressed in previous work.

\begin{figure}[bt!]
  \center
  \includegraphics[width=3.0in]{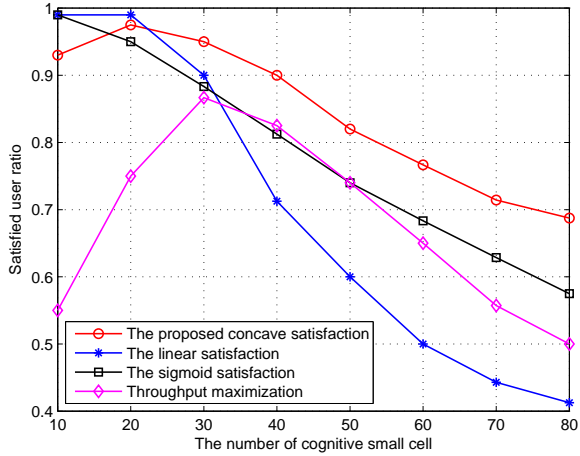}\\
  \caption{The comparison results of satisfied user ratio for different satisfaction utility functions.}
  \label{fig:satisfaction_comparison}
\end{figure}

To improve the network satisfaction, an efficient approach is to prevent the users compete for extra resources when it is satisfied and to decrease the satisfaction utility when it occupies additional resources. Based on this intuition, the concave satisfaction utilities may be more suitable for multiuser communication networks. An example of concave satisfaction utilities is given by
$\big(\frac{2\sqrt{r_nd_n}}{d_n+r_n}\big)^\alpha$, where $\alpha$ is used to adjust the slope of the  utility curve for different types of traffic. For illustration, we consider the problem of distributed spectrum access for CSCs. Specifically, the CSCs are randomly located in a region of 100m $\times$ 100m, and the sensing-based spectrum access protocol proposed in \cite{small_cell2} was applied. The problem of distributed spectrum access is formulated as a graphical game and the learning automata \cite{SLA} is applied. Different typical applications, such as G.711PCM, WMV, AVI/RM, Flash, H.264, are considered in the simulation. The comparison results are presented in Fig. \ref{fig:satisfaction_comparison}.
It is noted that with the proposed satisfaction function, the satisfaction user  ratio is largely improved. In particular, as the network scales up, the throughput gain becomes significant.

 \begin{figure}[bt!]
  \center
  \includegraphics[width=3.5in]{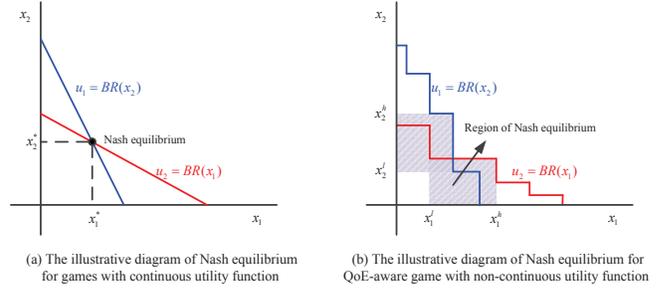}\\
  \caption{ An illustrative expansion of Nash equilibrium of discrete-QoE-aware game. $BR(x_2)$ ($BR(x_1)$) denotes  the best response curve of player 1 (player 2) with respect to the decision variable of the other player. In (a), the intersection point $(x^*_1, x^*_2,)$ is Nash equilibrium. In (b), due to the discontinuous feature of QoE-aware game, the Nash equilibrium is expanded to the shadow region.}
  \label{fig:QoE_game}
\end{figure}

\emph{2) Discrete-QoE-aware game:} Eventually, the purpose of wireless communications is to serve  people. Thus, the perception by people, i.e.,  QoE, should be included in the game formulation. Unlike the satisfaction function which is characterized by continuous and real values, the perception of people is generally subjective and discrete. For example, a person may feel  ``Excellent",  ``Good", ``Fair", ``Poor" and ``Bad", by the method of the mean opinion score \cite{QoE}.  Compared with traditional continuous optimization game, an interesting result of discrete-QoE-aware game is the expansion of Nash equilibrium, which is shown and illustrated in Fig. \ref{fig:QoE_game}.

In traditional continuous optimization game, users  maximize their throughput as there is an inherent principle: larger throughput is always better. On the contrary, a user in discrete-QoE-aware games does not always maximize its throughput unless its QoE level can be improved, e.g., from ``Poor" to ``Fair". Thus, it can be expected that the discrete-QoE-aware game would improve the network QoE.

To further show the benefit of discrete-QoE-aware game, we consider the problem of distributed user association in LTE-A small cell networks. For users located in the overlapping areas, there are multiple small cell access points (SAPs) available and the users need to choose one to associate. Consider three types of video calling users using Skype: i) the first one is the group video calling with the required minimal throughput $R_m=512$kbps and the recommended throughput $R_c=2$Mbps, ii) the second one is the high definition video calling  with $R_m=1.2$Mbps and $R_c=1.5$Mbps, and iii) the third one is the general video calling user with the
with $R_m=128$kbps and $R_c=500$kbps. Each user falls into one of the above three types with equal probabilities. It is believed that the minimal throughput only supports the basic user demand (``Poor"), while the recommended throughput offer sufficiently good user experience (``Good"). With the method proposed in \cite{MOS_classification},  the  throughput thresholds for other QoE levels (``Excellent",  ``Fair" and ``Bad") can be obtained accordingly.

  \begin{figure}[bt!]
  \center
  \includegraphics[width=3.0in]{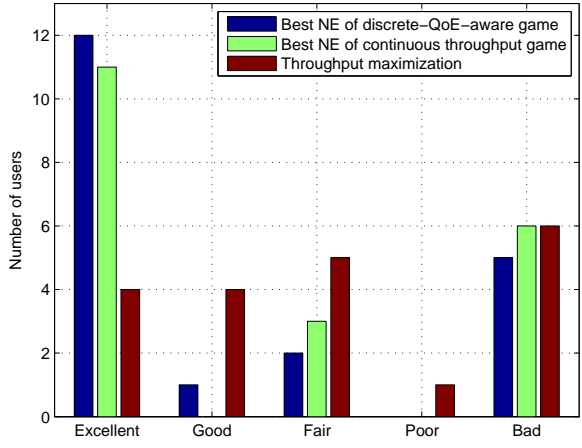}\\
  \caption{The number of users in different QoE levels of different  solutions. }
  \label{fig:QoE_comparison}
\end{figure}

Considering a network with 78 users which can access only one SAP, and 20 users which are located in the overlapping regions of neighboring CSCs, the comparison results of the number of users in different QoE levels are shown in Fig. \ref{fig:QoE_comparison}.
It is seen that the discrete-QoE-aware game outperforms the continuous optimization game. Specifically, with the discrete-QoE-aware game, 12 users are in ``Excellent", one  in ``Good", two  in ``Fair", and  five in ``Bad"; with the continuous optimization game, 11 users are in ``Excellent", none in ``Good",  three  in ``Fair" and  five  in ``Bad". Also, it is noted that the  throughput maximization approach (the user demand is neglected) achieves poor network QoE. This result validates the superiority of the discrete-QoE-aware game.

\emph{3) Hierarchical game:} As stated before, CSCs would interact with macro-cells for dynamic spectrum sharing and mobile offloading.
While originally studied in the economic context of duopolies in which one company has the power to act before the other companies, Stackelberg game, which is an important kind of hierarchical games, is suitable for systems that contain a natural hierarchy. Therefore, to address the hierarchical decision-making between macro-cells (always acting as a leader) and CSCs (always acting as followers), Stackelberg game is becoming a useful tool \cite{Stackelberg_game}.

\begin{figure}[bt!]
  \center
  \includegraphics[width=3.5in]{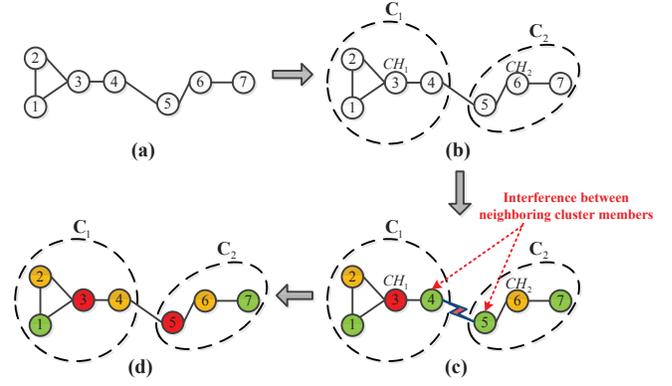}\\
  \caption{ An example of using ``divide-and-conquer" to use cluster-based hierarchical game in large-scale CSCs. In (c)  and (d), the  colors represent the channels chosen by the small cells.}
  \label{fig:stackelberg_game}
\end{figure}

In addition, following the idea of ``divide-and-conquer", hierarchical game can also be used to address the dense deployment of CSCs. In particular, in order to ease the challenges  caused by the large number of participants, we can create hierarchy to transform the large-scale optimization problem into several layers of sequential sub-problems. To achieve this, a useful method is clustering. An example of creating hierarchy to use cluster-based hierarchical game in large-scale CSCs is shown in Fig. \ref{fig:stackelberg_game}. In (a), the network topology and the interference relationship is presented. In (b), the neighboring small cells distributively  form two disjoint clusters  with cells 3 and cell serving as the  headers respectively. In (c), in the upper layer, the  headers compete for resources with each other, aiming to maximize the aggregate utility of the cluster; in the lower layer, the cluster members compete with other members, under the policies imposed by the header. In (d), since different clusters behave independently, there may be interference between neighboring clusters, e.g., cells 4 and 5 still interfere with each other. Thus, the interfering cells further mitigate mutual interference via distributed learning, e.g., Q-learning.
With the proposed cluster-based hierarchical structure, the self-organizing optimization in large-scale networks can then be solved with moderate computational complexity.

 \begin{figure}[bt!]
  \center
  \includegraphics[width=3in]{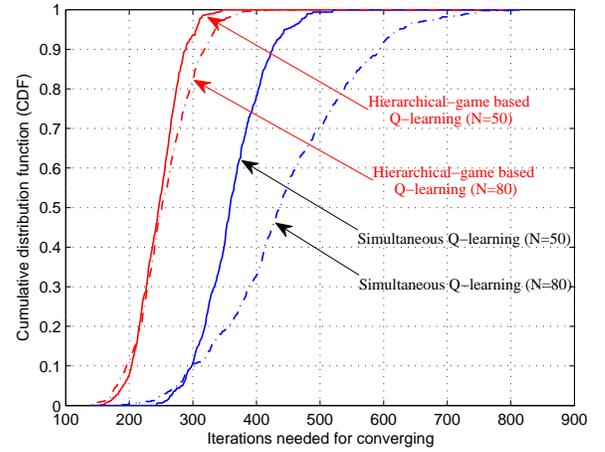}\\
  \caption{ The convergence speed comparison between the  hierarchical-game based Q-learning and simultaneous Q-learning.}
  \label{fig:hierarchical_comparison}
\end{figure}

We compare the computational complexity between the proposed cluster-based hierarchical game and the simultaneous Q-learning approach \cite{Independent_Q_learning}, in which all cells performs Q-learning simultaneously. The achievable network throughput of two approaches is almost the same.  The cumulative distribution function (CDF) of the iterations needed to converge is shown in
Fig. \ref{fig:hierarchical_comparison}. It is noted from the figure that for the same size network, e.g., $N = 50$ or $N = 80$, the iterations needed for converging of the  hierarchical-game Q-learning approach are significantly decreased. Furthermore, when the network scales up from $N = 50$ to $N = 80$, the convergence speed of the hierarchical-game based Q-learning approach slightly decreased while that of the simultaneous Q-learning approach  is largely decreased. This implies that the proposed hierarchical-game based approach  is especially suitable for dense and large-scale networks.

\emph{4) Robust game:} To capture the random and dynamic behaviors in CSCs,  robust game is a good candidate. Specifically, the utility function in robust games is defined over statistics \cite{robust_game}, e.g., expectation or other high-order statistics. In the following, we present a robust spectrum access game for CSCs   as an illustrative example.

Consider a distributed CSC network operating in the TV white space. Each cognitive SAP inquires the spectrum availability from the geo-location  database, which specifies the available channel set and the maximum allowable transmission power for each cell. To capture the dynamic cell load  in practical applications, we consider a network with a varying number of active cells. Specifically, it is assumed that each cell performs spectrum access with probability $\lambda_n$, $0 < \lambda_n \le 1$, in each decision period. Note that such a  model captures general kinds of dynamics in wireless networks, e.g., a cell becomes active only when it has data to transmit and inactive when there is no transmission demand. Also, it can be regarded as an abstraction of the dynamic cell loading, i.e., the cell active probability corresponds to the probability of a non-empty loading buffer. Note that the active cell set in each period is not deterministic and randomly changes from period to period. Also, a cell does not know the active probabilities of other cells.

  \begin{figure}[bt!]
  \center
  \includegraphics[width=3.0in]{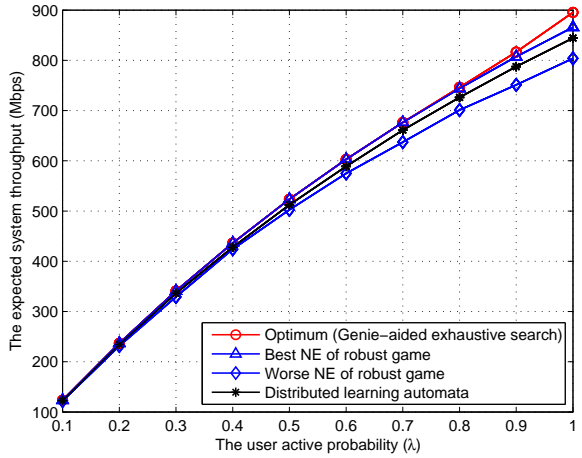}\\
  \caption{The expected Shannon capacity when varying the active probabilities of the cells. }
  \label{fig:robust_game}
\end{figure}

To address the dynamic and random deployment of CSCs, a robust spectrum access game can be formulated, in which the utility function of a CSC is defined as the expected Shannon capacity over all possible active cell sets. The game can be proved to be a potential game \cite{Monderer96} and hence the distributed learning automata algorithm \cite{SLA} can be applied to converge to NE points in the dynamic environment. Taking a network with nine CSCs as an illustrative example, the throughput performance comparison results are shown in Fig. \ref{fig:robust_game}. The optimum is obtained using the exhaustive search method in a centralized manner, by assuming that there is a genie that knows all required information. The best and worst NE is obtained using the best response algorithm in a distributed manner, by assuming that information exchange among neighboring cells is available. Some important results can be observed: i) the best NE is almost the same with the optimal one, while the throughput gap between the worse NE and the optimum is also trivial, which validates the effectiveness of the formulated robust spectrum game, ii) the achievable throughput of the distributed learning automata is very close to the optimal one.

\emph{5) Content-aware game: }As legacy cellular systems have migrated towards data and Internet services, taking into account the access content in the self-organizing optimization would enjoy content gain. For Internet traffic, it has been shown in \cite{content_classification} that a relatively small proportion of the access items accounts for a vast fraction of the information accesses, and the Zipf's law can be used to determine the occurrence frequency of the access items, given the content rank, the content pool size, and the characteristic curve of the access pattern. Nowadays, content caching has become a core technology for wireless cellular systems. Thus, it is reasonable to replicate significant portions of popular contents on the wireless caches. As a result, the search and access time for popular content is fast compared with that of unpopular content. The reason is that the popular content can be accessed in the wireless caches, while the unpopular content is accessed in the far-side server. Therefore, the differences in access time of different contents will have great impact on wireless resource allocation and  it is promising to explore content-aware game-theoretic solutions for CSCs, which would achieve better performance.

\vspace{-0.12in}

\subsection{Comparative summarization and analysis }
 In comparison, the  game-theoretic models for CSCs presented in this article different from previous ones significantly. Specifically, it shifts from throughput-oriented optimization to user-centric optimization, e.g., demand-aware game, discrete-QoE-aware game and content-aware game,  addresses the dense deployment of small cells, e.g., graphical game and hierarchical game, and copes with randomness and dynamics well, e.g., robust game. Although the research on game-theoretic self-organizing optimization for CSCs is in infancy,  we believe that the presented game-theoretic models will draw great attention in the near future.

For a specific resource optimization problem in CSCs, one can choose a suitable game-theoretic model and a learning algorithm to construct a self-organizing optimization solution. However, it should be pointed out that game-theoretic solution for CSCs is application-dependent, which means that the  game-theoretic model and distributed learning algorithm should be carefully formulated and designed.

 \subsection{Future research direction}
 It is seen that game-theoretic solutions for self-organizing optimization in CSCs have definitely drawn a beautiful and exciting future, though the current research  is still far away from the expected vision. We list some  future research problems for game-theoretic models and learning procedures below:
\begin{enumerate}
  \item Develop or investigate new game-theoretic models for self-organizing optimization from social/biological behaviors. The  rationale behind is that in old days humanity first self-organized and then evolved successfully with population growth. For example, motivated by the local altruistic behavior in biological systems,  a local altruistic game with each player maximizing its utility and the aggregate utilities of its neighbors was proposed to achieve global optimization via local information exchange \cite{graphical_game}. The key design of this issue is to properly abstract and model the social/biological behaviors, which is interesting and challenging.
  \item It is noted that each kind of presented game mainly addresses a single aspect of challenges in CSCs. However, as can be expected, one may combine more than one game-theoretic model, e.g., robust discrete-QoE-aware game or Stackelberg graphical game, to address multiple aspects of challenges of CSCs simultaneously. Such combinations bring about new challenges since the game structure is completely changed.
   \item Design and analyze heterogeneous learning algorithms. In most existing studies, it is assumed that all the decision-makers employed the same learning algorithm. However, this assumption is  for academic research but not true in practical systems. In practice,  the small cells may belong to different holders, which may adopt different learning algorithms; in addition, even the small cells belonging to the same holder may have different processing ability and preference, and hence choose heterogeneous learning algorithms. Introducing heterogeneity into the learning procedure will change the convergence and asymptotic behavior, which needs to be further studied.
   \item Design knowledge-assisted learning algorithms. The common  procedure in existing learning algorithms is to update the strategies based on the historical action-payoff information. It may take long time to converge to stable solutions since the players need to explore all the possible actions. As shown in Fig. \ref{fig:paradgim}, knowledge can be viewed as high-level intelligence obtained from the contextual  information, which is truly beneficial for decision-making. Thus, we should develop some new knowledge-assisted learning technologies to increase the converging speed and achieve better performance.
\end{enumerate}

 \section{Conclusion}
In this article, we investigated self-organizing optimization for CSCs, which will play an important role in future cognitive cellular systems. By exploring the inherent features, some fundamental challenges  and requirements for self-organizing optimization in CSCs were presented and discussed. Following the attractive advantages of game-theoretic models, i.e., distributed and autonomous decision-making,  a framework of game-theoretic solutions for self-organizing optimization in CSCs was established, and some featured game-theoretic models were proposed. Specifically, the basic game-theoretic models are presented,  some insights are discussed, some examples are discussed and  future research directions are given.

\ifCLASSOPTIONcaptionsoff
  \newpage
\fi

\bibliographystyle{IEEEtran}
\bibliography{IEEEabrv,reference}
\end{document}